\documentclass[11pt]{article}
\usepackage{setspace}
\usepackage{amsfonts}
\usepackage{graphicx,subfig}
\usepackage{amsmath,amsthm}
\usepackage{wrapfig}

\newtheorem{thm}{Theorem}[section]
\newtheorem{lem}{Lemma} [section]
\newtheorem{definition}{Definition}
\newtheorem{cor}{Corollary}
\newtheorem{alg}{Algorithm}
\newtheorem{remark}{Remark}

\newcommand{\Cr}{\mathcal{C}r}
\newcommand{\A}{{\bf A}}

\newcommand{\E}{{\bf E}}
\def\x{{\bf x}}
\def\y{{\bf y}}
\def\w{{\bf w}}
\def\v{{\bf v}}
\def\u{{\bf u}}
\def\z{{\bf z}}
\def\e{{\bf e}}
\def\ddelta{\bf{\delta}}

\newcommand{\beq}{\begin{equation}}
\newcommand{\eeq}{\end{equation}}
\newcommand{\bea}{\begin{eqnarray}}
\newcommand{\eea}{\end{eqnarray}}

\newcommand{\stexp}{\mbox{$\mathbb{E}$}}    
\newcommand{\Prob}{\ensuremath{\mathbb{P}}}

\long\def\symbolfootnote[#1]#2{\begingroup%
\def\thefootnote{\fnsymbol{footnote}}\footnote[#1]{#2}\endgroup}

\begin{document}

\title{Sparse Recovery of Positive Signals with Minimal Expansion }
\author{ M.Amin Khajehnejad \, Alexandros G. Dimakis \, Weiyu Xu \,
  Babak Hassibi \\ California Institute of Technology \\{\tt email: \{amin,adim,weiyu,hassibi\}@caltech.edu}}
 \maketitle

\vspace{1cm}

\begin{singlespace}
\begin{abstract}
We investigate the sparse recovery problem of reconstructing a
high-dimensional non-negative sparse vector from lower dimensional
linear measurements. While much work has focused on dense measurement
matrices, sparse measurement schemes are crucial
in applications, such as DNA microarrays and sensor networks, where
dense measurements are not practically feasible. One possible
construction uses the adjacency matrices of expander graphs, which
often leads to recovery algorithms much more efficient than $\ell_1$
minimization. However, to date, constructions based on expanders have
required very high expansion coefficients which can potentially make
the construction of such graphs difficult and the size of the
recoverable sets small.

In this paper, we construct sparse measurement matrices for the
recovery of non-negative vectors, using perturbations of the adjacency
matrix of an expander graph with much smaller expansion
coefficient. We present a necessary and sufficient condition for
$\ell_1$ optimization to successfully recover the unknown vector and
obtain expressions for the recovery threshold. For certain classes of
measurement matrices, this necessary and
sufficient condition is further equivalent to the existence of a
``unique'' vector in the constraint set, which opens the door to alternative
algorithms to $\ell_1$ minimization. We further show that
the minimal expansion we use is necessary for any graph for which
sparse recovery is possible and that therefore our construction is
tight. We finally present a novel recovery algorithm that exploits
expansion and is much faster than $\ell_1$ optimization. Finally, we
demonstrate through theoretical bounds, as
well as simulation, that our method is robust to noise and approximate
sparsity.
\end{abstract}
\end{singlespace}

\thispagestyle{empty} 


\newpage
\setcounter{page}{1}

\section{Introduction}

We investigate the problem of signal recovery in compressed sensing,
i.e., the problem of reconstructing a signal $\x$ that is assumed to
be $k$ sparse using $m$ measurements, $\y=\A\x$, where $m$ is smaller than the
ambient dimension of the signal $n$, but larger than $k$. $\A$
here is the $m\times n$ so-called measurement matrix. In this
paper, we focus on the case where the nonzero entries of $\x$ are
positive, a special case that is of great practical interest.

In compressed sensing, $\A$ is often a dense matrix drawn from some
ensemble of random matrices (see, e.g., \cite{Candes and Tao general
  case}). In this paper, however, we will focus on sparse measurement
matrices. This is important for numerous reasons.
In several applications, like DNA micro arrays, the cost of each
measurement increases with the
number of coordinates of $\x$ involved~\cite{DNA microarray,Dai_DNA}.
Also, sparse measurement matrices often make possible the design of
faster decoding algorithms (e.g., \cite{Weiyu expander
  algorithm,Indyk_Ruzic,Tropp_OMP,Guru2}) apart from the general linear
programming decoder \cite{Candes and Tao general case}.
In addition, unlike random measurement matrices (such as
Gaussian or Bernoulli), which only guarantee the recovery of sparse
vectors with high probability, expander graphs give deterministic
guarantees (see, e.g., \cite{Weiyu expander algorithm}, which gives a
deterministic guarantee for the fast algorithm proposed, and
\cite{Indyk RIP} for concentration lemmas on expander graphs).

Unlike Gaussian matrices, where reasonably sharp bounds on the
thresholds which guarantee linear programming to recover sparse
signals have been obtained \cite{Neighborliness of Gaussians}, such
sharp bounds do not exist for
expander-graph-based measurements. This is the main focus of the
current paper, for the special case where the $k$-sparse vector is
non-negative.

It turns out that, due to the additional non-negativity constraint,
one requires significantly fewer measurements to recover $k$-sparse
non-negative signals. The non-negative case has also been studied in
\cite{Donoho Tanner nonnegative} for Gaussian matrices and also in the
work of Bruckstein et al.~\cite{BRZ_nonneg}, which further proposes a
``matching pursuit'' type of recovery algorithm. See also~\cite{Dai_Positive} for another example.

The success of a measurement matrix is often certified by a
so-called Restricted Isometry
Property (RIP) which guarantees the success of $\ell_1$ minimization.
Recently, Berinde et al.~\cite{Indyk RIP} showed that the adjacency
matrices of suitable unbalanced expander graphs satisfy an RIP
property for $\ell_{p \sim 1}$ norm.
However, it turns out that RIP conditions are only sufficient and
often fail to characterize all good measurement matrices.
A complete characterization of good measurement matrices was recently
given in terms of their
null space. More precisely, as stated in previous work
(e.g. ~\cite{Weiyu Mihailo Null Space,Null space proof alternative
  1,Null space proof alternative 2,Null space proof alternative 3}),if
for any vector $\w$ in the null
space of $\A$, the sum of the absolute values of any $k$ elements of $\w$
is less that the sum of the absolute values of the rest of the elements,
then the solution to $\min{\|\x\|_0} \: \text{subject to} \: \A \x=\y$ can
always be obtained by solving $\min{\|\x\|_1} \: \text{subject to} \:
\A\x=\y$, provided $\x$ is $k$-sparse.\footnote{Here $\|\cdot\|_0$
represents the number of non-zero entries in its argument vector and
$\|\cdot\|_1$ is the standard $\ell_1$-norm.} This condition is stated
in the work of Donoho~\cite{Donoho Neighborliness} as the $k$-neighborly
polytope property of $\A$,  and in the work of Candes et~al.~as the
uncertainty principle~\cite{Candes and Tao general case}. Donoho
et~al.~also have been able to show the
validity of this condition with high probability for random
i.i.d Gaussian matrices and are therefore able to compute fairly tight
thresholds on when linear-programming-based compressed sensing works
\cite{Neighborliness of Gaussians}.
The first analysis of the null space for expander graphs has been done
by Indyk~\cite{Indyk nullspace}, where it was shown that every
($2k$,$\epsilon$) expander graph\footnote{We shall formally define
  $(k,\epsilon)$ expander graphs shrotly.} with $\epsilon\leq\frac{1}{6}$ will
have a well supported null space. See also~\cite{Guru1} for explicit constructions using expander graphs.

%
%
Furthermore, using Theorem 1 of \cite{Weiyu  Grassman manifold}, which is a
generalization of the null space property theorem for the recovery of
approximately sparse signals, Indyk's result gives an upper bound on
the error when linear programming is used to
recover approximately $k$-sparse vectors using expander graph
measurements.

\noindent\textbf{Contributions of the current paper.}
We present a necessary and sufficient condition that completely
characterizes the success of $\ell_1$-minimization \emph{for
  non-negative signals}, similar to the null space condition for the
general case. Our condition requires that all the vectors in the null
space of $\A$ have sufficiently large \emph{negative support} (i.e. a
large number of negative entries). It further turns out that, for a
certain class of measurement matrices $\A$, this
condition is nothing but the condition for the existence of a ``unique''
vector in the constraint set $\{\x|\A\x=\y,\x\geq 0\}$. This therefore
suggests that any other convex optimization problem over this
constraint set can find the solution. (We exploit this fact later to
find faster alternatives to $\ell_1$ minimization.)

We then use the necessary and sufficient characterization to construct
sparse measurement matrices. Our construction relies on starting with the
adjacency matrix of an unbalanced expander (with constant degree) and
adding suitable small perturbations to the non-zero entries.

Several sparse matrix constructions rely on adjacency matrices
of expander graphs~\cite{Indyk RIP,Indyk nullspace,Weiyu expander
  algorithm,Weiyu_Exp2,Indyk Explicit construction}.
In these works, the technical arguments require very large expansion
coefficients, in particular, $1-\epsilon \geq 3/4$, in order to guarantee a
large number of unique neighbors \cite{expcodes} to the expanding
sets. A critical innovation of our work is that we require much
smaller expansion, namely, $1-\epsilon=1/d$, where
$d$ is the number of non-zero entries in every column of $\A$. In fact,
we show that expansion of $1-\epsilon=\frac{1}{d}$ is \emph{necessary}
for any matrix that works for compressed sensing.
These two results show that for nonnegative vectors, expansion of
$1-\epsilon=\frac{1}{d}$ is necessary and sufficient, and the small
expansion requirement allows a much larger set of recoverable
signals.

The reason for this different requirement is that we use expansion in
a different way than previous work; we do not require a unique
neighbor property but rather rely on Hall's theorem that guarantees
that $1-\epsilon=1/d$ expansion will guarantee perfect matchings for
expanding sets. The matching combined with perturbations in the
entries guarantees full rank sub-matrices which in turn translates to
the null space characterization we need.

Finally, we propose a fast alternative to $\ell_1$ optimization for
recovering the unknown $\x$. The method first identifies a large
portion of the unknown vector $\x$ where the entries are zero, and
then solves an ``overdetermined'' system of linear equations to
determine the remaining unknown components of $\x$. Simulations are
given to present the efficacy of the method and its robustness to
noise.

\section{Problem Formulation}
\label{sec:problem}
The goal in compressed sensing is to recover a sparse vector from a set of
under-determined linear equations. In many real world
applications the original data vector is nonnegative, which is the
case that we will focus on in this paper.   The original
problem of compressed sensing for the nonnegative input vectors is the
following:
\begin{equation} \label{NP}
\min_{\A \x=\y,\x\geq 0}{\|\x\|_0}
\end{equation}
where $\A^{m\times n}$ is the measurement matrix, $y^{m\times 1}$ is
called the observation vector, $x^{n\times 1}$ is the unknown vector
which is known to be $k$-sparse, i.e., to have only $k$ nonzero
entries, and where $\|\cdot\|_0$ is $l_0$ norm, i.e., the number of
nonzero entries of a given vector. The typical situation is that
$n>m>k$. Although (\ref{NP}) is an NP-hard
problem, Donoho and Tanner have shown in \cite{Donoho Tanner
  nonnegative}
that, for a class of  matrices $\A$ maintaining a so-called outwardly
$k$-neighborly property and  $\x$ being at most $k$-sparse, the solution to
(\ref{NP}) is unique and  can be recovered via the following linear
programming problem:
\begin{equation} \label{LP}
\min_{\A \x=\y,\x\geq0}{\|\x\|_1}
\end{equation}
They also show that i.i.d Gaussian random  $m\times n$
matrices with $m=n/2$ are outwardly $m/8$-neighborly with high
probability, and thus allow the recovery of  $n/16$ sparse vectors
$\x$ via linear programming. They further define a \emph{weak}
neighborly notion, based upon which they show that the same Gaussian
random matrices will allow the recovery of
\emph{almost} all $0.279n$ sparse vectors $\x$ via $\ell_1$-optimization
for sufficiently large $n$.

In this paper, we primarily seek the answer to a similar question
when the measurement matrix $\A$ is sparse and, in particular when
$\A$ is the adjacency matrix of
an unbalanced bipartite graph with constant left degree $d$.
The aim is to analyze the
outwardly neighborly conditions for this class of matrices and come
up with sparse structures that allow the recovery of vectors with sparsity
proportional to the number of equations.


\section{Null Space Characterization and Complete Rank}\label{Sec:
  Completer Rank}
\label{sec:nullspace}

We begin by stating an equivalent version of the outwardly neighborly
condition which is in fact similar to the null space property that was
mentioned in the introduction, but for the non-negative case. Later we
show that this has a much more mundane interpretation for the special
case of regular bipartite graphs. We present the first theorem in the
same style as in \cite{Donoho Tanner nonnegative}.

\begin{thm}\label{Null Space}
Let $\A$ be a nonnegative $m\times n$ matrix and $k<n/2$ be a positive
integer. The following two statements are equivalent:
\begin{enumerate}
\item For every nonnegative vector $\x_0$ with at most $k$ nonzero
  entries, $\x_0$ is the unique solution to (\ref{LP}) with $\y=\A \x_0$.
\item For every vector $\w\neq 0$ in the null space of $\A$, and every index
set $S\subset\{1,2,...,n\}$ with $|S|=k$ such that $\w_{S^c}\geq 0$,
it holds that
\[ \sum_{i=1}^{n}w_i>0. \]
\end{enumerate}
Here $S^c$ is the complement set of $S$ in $\{1,2,...,n\}$ and
$\w_{S}$ denotes the sub-vector of $\w$ constructed by those elements
indexed in $S$. $|S|$ means the cardinality of the the set $S$
\end{thm}

Theorem \ref{Null Space} is in fact the counterpart of Theorem 1 of
\cite{Weiyu Mihailo Null Space} for nonnegative vectors. It gives a
necessary and sufficient condition on the matrix $\A$, such that all $k$-sparse
vectors $\x_0$ can be recovered using (\ref{LP}). The condition is essentially
that for every vector in the null space of $\A$,
the sum of every $n-k$ nonnegative elements should be greater than the
absolute sum of the rest. (This is very similar, but not identical, to the
null space property of \cite{Weiyu Mihailo Null Space}.) Therefore we call
it the non-negative null space property.

\begin{proof}
Suppose $\A$ has the non-negative null space property. We assume $\x_0$ is
$k$-sparse and show that under the mentioned null space condition, the
solution to (\ref{LP}) produces $\x_0$. We denote by $\x_1$ the solution to
(\ref{LP}). Let S be the support set of $\x_0$. We can write:
\vspace*{-2pt}
\begin{eqnarray}
\nonumber \|\x_1\|_1&=&\|\x_0+(\x_1-\x_0)\|_1 \\
&=& \sum_{i=1}^{n}{\x_0(i) + (\x_1-\x_0)(i)} \label{norm inequalitites1} \\
&=& \|\x_0\|_1 +  \sum_{i=1}^{n}{(\x_1-\x_0)(i)} \label{norm inequalitites2}
\end{eqnarray}

\noindent  Where $\x_0(i)$ and $(\x_1-\x0)(i)$ are the $ith$ entry of $\x_0$ and $\x_1-\x_0$ respectively. The reason (\ref{norm inequalitites1}) and (\ref{norm inequalitites2}) are true is that $\x_1$ and $\x_0$ are both nonnegative vectors and their $\ell_1$-norm is simply the sum of their entries. Now, if $\x_1$ and $\x_0$ are not equal, since $\x_1-\x_0$ is in the null
space of $\A$ and is non-negative on $S_c$ (because $S$ is the support set of $\x_0$) we can write:
\begin{equation}
 \sum_{i=1}^{n}{(\x_1-\x_0)(i)} > 0
\end{equation}
which implies
\begin{equation}
\nonumber \|\x_1\|_1 > \|\x_0\|_1
\end{equation}
But we know that $\|\x_1\|_1\leq \|\x_0\|_1$ from the construction. This means that we should have $\x_1=\x_0$.

\noindent Conversely, suppose there is a vector $\w$ in the null space of $\A$ and a
subset $S\subset\{1,2,...,n\}$ of size $k$ with $\w_{S^c}\geq0$ and
$\sum_{i=1}^{n} w_i \leq 0$. We construct a non-negative vector $\x_0$ supported on $S$, and show that there exist another nonnegative vector $\x_1\neq \x_0$ such that $\A \x_0 = \A \x_1$ and $\|\x_1\| \leq \|\x_0\|$. This means that $\x_0$ is not the unique solution to (\ref{LP}) with $\y=\A \x_0$ and will complete the proof. For simplicity we may assume
$S=\{1,2,...,k\}$. We construct a nonnegative vector $\x_0$ supported on $S$ that cannot be recovered via $\ell_1$-minimization of (\ref{LP}).
Wlog we write
{\small
\begin{equation}
\w=\left[\begin{array}{c} -\w_{S^-} \\ \w_{S^+} \\ \w_{S^c}
\end{array}\right],
\end{equation}
}
where $\w_{S^-}$ and $\w_{S^+}$ are both nonnegative vectors. Now set
{\small
\begin{equation}
\x_0=\left[\begin{array}{c} \w_{S^-} \\ 0  \\ 0
\end{array}\right] ,~ \x_1=\left[\begin{array}{c} 0\\  \w_{S^+} \\ \w_{S^c}
\end{array}\right].
\end{equation}
}
\end{proof}

In this paper we will be considering measurement matrices $\A$ which
possess the following two features: 1.~the entries are non-negative
and 2.~the sum of the columns are constant. This class of matrices
includes measurement matrices obtained from the adjacency matrix of
regular left degree bipartite graphs (which have a constant number of
ones in each column), as well as their perturbed versions introduced
in section \ref{sec: perturbed exp.}. For this class of matrices we actually show that the
condition for the success of $\ell_1$ recovery is simply the condition
for there being a ``unique'' vector in the constraint set $\{\x|\A\x =
\A\x_0,\x\geq 0\}$. To this end, we prove the  following theorem

\begin{thm} Let $A\in{\cal R}^{m\times n}$ be a matrix with
  non-negative entries and constant column sum. Then the following three
  statements are equivalent.
\begin{enumerate}
\item For all non-negative $k$-sparse $\x_0$, it holds that
\[ \{\x|\A\x = \A\x_0,\x\geq 0\} = \{\x_0\}. \]

\item For every vector $\w\neq 0$ in the null space of $\A$, and every index
set $S\subset\{1,2,...,n\}$ with $|S|=k$ such that $\w_{S^c}\geq 0$,
it holds that
\[ \sum_{i=1}^{n}w_i>0. \]

\item For every subset $S\subset\{1,2,...,n\}$ with $|S|=k$,  there exists no vector $\w\neq 0$ in the null space of $\A$ such that  $\w_{S^c}\geq0$.
\end{enumerate}
\label{thm:unique}
\end{thm}


Theorems \ref{Null Space} and \ref{thm:unique} show that for the class of matrices with non-negative entries and constant column sum, the
condition for the success of $\ell_1$ recovery is simply the condition
for there being a ``unique'' vector in the constraint set $\{\x|\A\x =
\A\x_0,\x\geq 0\}$. In this case, {\em any} optimization problem,
e.g., $\min_{\x\geq 0,\A\x=\y}\|\x\|_2$, would also recover the desired
$\x_0$.

In fact, rather than prove Theorem \ref{thm:unique}, we shall prove
the following stronger result (from which Theorem \ref{thm:unique}
readily follows).

\begin{lem} Let $A\in{\cal R}^{m\times n}$ be a matrix with
  non-negative entries and constant column sum. Then the following two
  statements are equivalent.
\begin{enumerate}
\item For all non-negative $\x_0$ whose support is $S$, it holds that
\[ \{\x|\A\x = \A\x_0,\x\geq 0\} = \{\x_0\}. \]
\item For every vector $\w\neq 0$ in the null space of $\A$ such that $\w_{S^c}\geq 0$,
it holds that
\[ \sum_{i=1}^{n}w_i>0. \]

\item There exists no vector $\w\neq 0$ in the null space of $\A$ such that  $\w_{S^c}\geq0$.
\end{enumerate}
\label{lem:unique}
\end{lem}

\begin{proof}
First, we show that for any nonnegative matrix $\A$, statements $1$ and $3$ of Lemma \ref{lem:unique} is equivalent. suppose that the condition $3$ holds for a specific subset $S\subset\{1,2,\cdots,n\}$. Consider a nonnegative $n\times 1$ vector $\x_0$ supported on $S$. If there exists another nonnegative vector $\x_1$ with the property that $\A\x_1=\A\x_0$, then  $\x_1-\x_0$ would be a vector in the null space of $\A$ which is also nonnegative on $S^c$, due to the nonnegativity of $\x_1$ and the fact that $S$ is the support set of $\x_0$. This contradicts the earlier assumption of condition $2$.

The proof of converse is also straight forward. Suppose the condition $1$ holds for a specific subset $S$ and all nonnegative vectors $\x_0$ supported on $S$. Let's say one can find a nonzero vector $\w$ in the null space of $\A$ with $\w_{S^c}\geq0$ . As in the proof of Theorem \ref{Null Space}, we may write $\w$ as
{\small
\begin{equation}
\w=\left[\begin{array}{c} -\w_{S^-} \\ \w_{S^+} \\ \w_{S^c}
\end{array}\right]
\end{equation}
}
where $\w_{S^-}$ and $\w_{S^+}$ are both nonnegative vectors.  Now if
{\small
\begin{equation}
\x_0=\left[\begin{array}{c} \w_{S^-} \\ 0  \\ 0
\end{array}\right],~ \x_1=\left[\begin{array}{c} 0\\  \w_{S^+} \\ \w_{S^c}
\end{array}\right],
\end{equation}
}
\noindent then $\x_0$ and $\x_1$ are distinct nonzero vectors and belong to the set $\{\x|\A\x = \A\x_0,\x\geq 0\}$. This is a contradiction to the assumption we earlier made.
 
So far we have shown that for any nonnegative matrix $\A$ the two statements $1$ and $3$ are equivalent. Now we show that for matrices with constant column sum the two statements $2$ and $3$ are equivalent. We make use of Lemma \ref{lem:|w+|=|w-|} in Section \ref{sec: Null space of adj. matrix}, that for this special class of matrices with constant column sum, every vector in the null space has a zero sum of entries. Therefore, statement $2$ can be true only if there is no $\w$ in the null space of $\A$ with $\w_{S^c}\geq 0$. Conversely if the condition in statement $3$ holds, then there is no $\w\in \mathcal{N}(\A) \setminus\{0\}$ such that $\w_{S^c}$ is nonnegative and therefore statement $2$ is also true.
 
\end{proof}

\subsection{Null Space of Adjacency Matrices} \label{sec: Null space of adj. matrix}

As promised earlier, we will now assume that $\A$ is the adjacency
matrix of a bipartite graph with $n$ nodes on the left and $m$ nodes
on the right. We also assume that the graph is left
$d$-regular. In other words $\A$ is a ($m\times n$) matrix with exactly $d$
ones in each column. We will now give a series of results for such
matrices. However, we should note that, unless stated otherwise, all
these results continue to hold for the class of matrices with
non-negative entries and constant column sum (the interested reader
should be able to easily verify this).

\begin{lem}\label{lem:|w+|=|w-|}
Let $\A^{m\times n}$ be a matrix with nonnegative entries and constant column sum.  For any vector $\w$ in the null space of $\A$, the following is true
\begin{equation}
\sum_{i=1}^{n}w_i = 0
\end{equation}
\end{lem}
\begin{proof} Let ${\mathbb{\bf 1}} =[1,1,...,1]$ be the $m\times 1$ vector of all $1$'s. We have:
\begin{equation}
\A\w=0\Rightarrow \mathbb{\bf 1}\A\w=0\Rightarrow d\sum_{i=1}^{n}\w_i=0
\end{equation}
\end{proof}

\begin{thm}\label{thm:null space for regular graphs}
For any matrix $\A^{m\times n}$  with exactly $d$ 1's in each column and other entries zero, the following two statements are equivalent:

$\bullet$ Every nonnegative vector $\x_0$ with at most $k$ nonzero entries is the unique solution to (\ref{LP}) with $\y=\A \x_0$.

$\bullet$ Every vector $\w$ in the null space of $\A$ has at least $k+1$ negative entries.
\end{thm}

\begin{proof}
We only need to show that for any $\w\in\mathcal{N}(\A)/\{0\}$ the second
statements of Theorem \ref{Null Space} and Theorem \ref{thm:null space
  for regular graphs} are equivalent. Let's assume there exists a nonzero
$\w\in \mathcal{N}(\A)$ with at most $k$ negative entries. We use
$S_{\w}^+$, $S_{\w}^-$ and $S_{\w}^0$ to denote the support of
positive, negative and zero entries of $\w$ respectively. By Lemma
 \ref{lem:|w+|=|w-|},
$\sum_{i=1}^{n}w_i=0$. Therefore if $S = S_{\w}^+\cup S_{\w}^0$, then $|S|\leq k$ and the nonnegative null space property is not satisfied (for the set $S$).

\noindent The other direction is straightforward. If any
$\w\in\mathcal{N}(\A)$ has $k+1$ negatives, there is no choice for
$S\subset \{1,2,...,n\}$ of size $k$, with $\w_{S^c} \geq 0$ and the nonnegative null space property is satisfied already.
\end{proof}

These results show how the structure of the null
space of the measurement matrix is related to the recoverability of
sparse vectors. Thus to achieve our primary goal,
which is constructing optimal sparse measurement matrices, we need to
find bipartite graphs with non-negative null space properties up to a
maximal sparsity (hopefully, proportional to the dimension $n$). One
promising choice would be the use of the adjacency matrix of expander
graphs. However, rather than restrict ourselves to this choice, we
present some theorems paraphrasing the
null-space property and interpreting it in terms of other properties
of matrices. This way we show that at some point
expander graphs inevitably emerge as the best choice, and even further as a necessary condition for any measurement matrix.

\subsection{Complete Rank and Natural Expansion} \label{sec: Cr and N. Exp.}

Before proceeding, let us consider the following two
definitions, whose relation to the main topic will be shortly made
apparent.
\begin{definition}
For a matrix $\A^{m\times n}$ we define the \emph{Complete Rank} of $\A$ (denoted by $\Cr(\A)$) to be the maximum integer $r_0$ with the property that every $r_0$ columns of $\A$ are linearly independent. In other words, $\Cr(\A)=\min_{\w\in\mathcal{N}(\A)}(|Supp(\w)|-1)$, where by $Supp(\w)$ we mean the support set of $\w$.
\end{definition}

This notion is also known in linear algebra as Kruskal rank (see \cite{Kruskal}).

\begin{definition}
A left regular bipartite graph ($X$,$Y$,$d$) with $X$ and $Y$ the set of left and right vertices ($|X|=n$,$|Y|=m$) and $d$ the regular left degree  is called a ($k$,$\epsilon$)-unbalanced expander if for every $S\subset X$ with $|S|\leq k$, the following holds: $|\Gamma(S)|\geq kd(1-\epsilon)$, where $\Gamma(S)$ is the set of neighbors of $S$ in $Y$.
\end{definition}

The following lemma connects these two notions:
\begin{lem}\label{CR and expanders}
Every bipartite graph with adjacency matrix $\A$ and left degree $d$ is a ($\Cr(\A)$,$\frac{d-1}{d}$) expander.
\end{lem}

\begin{proof}
If $S\subset X$ with $|S|\leq \Cr(\A)$ then the columns of $\A$
corresponding to the elements of $S$ are linearly independent. So the
sub-matrix of $\A$ produced by only those columns which correspond to
$S$ must be of full rank. Therefore, it must have at least $|S|$
non-zeros rows, which is to say $|\Gamma(S)|\geq
|S|=|S|d(1-\frac{d-1}{d})$.
\end{proof}

A direct corollary of this theorem is that:
\begin{equation}\label{eq:CR_Neighbors}
\forall S\subseteq X \mbox{, } |\Gamma(S)|\geq \min(|S|,\Cr(\A))
\end{equation}
The notion of complete rank is tightly
related to the expansion property. It is also related to the null
space characterization we are shooting for. The following theorem sheds some light on this fact.

\begin{thm}\label{thm:positive negative ratio}
If $\A^{m\times n}$ is the adjacency matrix of a left $d$-regular bipartite graph, then for every nonzero vector $\w$ in the null space of $\A$ the number of negative elements of $\w$ is at least $\frac{\Cr(\A)}{d}$.
\end{thm}

\begin{proof}
Let $X$ and $Y$ be the sets of left and right vertices of the
bipartite graph corresponding to $\A$ ($X$ corresponds to columns of
$\A$). let $S_{\w}^+$ be the set of vertices in $X$ corresponding to
the positive elements of $\w$, and likewise $S_{\w}^-$ be set of
vertices corresponding to the negative elements.\footnote{We
  interchangeably use $S$ and its variations to denote a set of
  vertices or a support set of a vector.} Let $S_{\w}=S_{\w}^+\cup
S_{\w}^-$. The equation $\A \x=\y$ can be manifested on the graph
representation of $\A$ with each node of $Y$ corresponding to an
equation with zero R.H.S.  This entails
$\Gamma(S_{\w}^+)=\Gamma(S_{\w}^-)=\Gamma(S_{\w})$, since otherwise,
there exists a vertex in $Y$ connected to exactly one of the sets $S_{\w}^+$ or
$S_{\w}^+$, and its corresponding equation will not sum up to zero. On
the other hand, from the definition of $\Cr(\A)$, we must have $|S_{\w}|\geq
\Cr(\A)$. The number of edges emanating from $S_{\w}^-$ is
$d|S_{\w}^-|$, which is at least as large as the number of its
neighbors $|\Gamma(S_{\w}^-)|$. Then:

\begin{eqnarray*}
 d|S_{\w}^-| \geq |\Gamma(S_{\w}^-)|=|\Gamma(S_{\w})|\geq\Cr(\A)
\end{eqnarray*}
Where the last inequality is a consequence of (\ref{eq:CR_Neighbors}).
\end{proof}

We now turn to the task of constructing adjacency matrices with
complete rank proportional to dimension. Throughout this paper, all
the thresholds that we achieve are asymptotic, i.e., they hold for the
regime of very large $n$ and $m$.

\subsection{Perturbed Expanders}\label{sec: perturbed exp.}

When $n$ and $m=\beta n$ are large, we are interested in constructing
0-1 matrices $\A^{m\times n}$ with  $d$ (constant) 1's in each column
such that $\Cr(\A)$ is proportional to $n$.  Furthermore, the
maximum achievable value of $\frac{\Cr(\A)}{nd}$ is critical. This is a very
difficult question to address. However, it turns out to be much easier
if we allow for a {\em small} perturbation of the nonzero entries of
$\A$, as shown next.

\begin{lem}\label{lem: perturbations}
For a matrix $\A\in \mathbb{R}^{m \times n}$ which is the adjacency
matrix of a bipartite left $d$-regular graph, if the submatrix formed by any
$r_0$ columns of $\A$ has at least $r_0$ nonzero rows, then it is possible
to perturb the nonzero entries of $\A$  and obtain another nonnegative  matrix $\tilde{\A}$ through this procedure, with
$\Cr(\tilde{\A})\geq r_0$. Furthermore,
the perturbations can be done in a way that the sum of each column remains
a constant $d$.
\end{lem}

\begin{proof}
The proof is based on showing that the set of valid perturbations that
do not guarantee $\Cr(\tilde{\A})\geq r_0$ is measure zero. So, by
choosing perturbations uniformly from the set of valid perturbations,
with probability one we will have $\Cr(A)\geq r_0$.
\end{proof}

\begin{remark}
It is worth mentioning that in a more practical scenario, what we
really need is that every $r_0$ columns of $\tilde{\A}=\A+\E$ be not
only nonsingular, but ``sufficiently'' nonsingular. In other words, if the
minimum singular value of the submatrix formed by any $r_0$ columns of
$\tilde{A}$ is greater than a constant number $c$ (which does not
depend on $n$), then that is what we recognize as the $RIP-2$ condition
for $\tilde{\A}$. This condition then guarantees that the solution to
$\ell_1$-minimization is robust, in an $\ell_2$-norm sense, to the
noise (please see \cite{Candes RIP noise} for more details on this
issue). Besides, in order to have a recovery algorithm which can be
implemented with reasonable precision, it is important that we have
such a well-conditioned statement on the measurement matrix. We will
not delve into the details of this issue here. However, we
conjecture that by leveraging ideas from perturbation theory, it
is possible to show that $\tilde{A}$ maintains some sort $RIP-2$
condition. Particularly, it has been shown in \cite{Perturbation
  Stewart} that random dense perturbations  $\tilde{\A} = \A + \E$
will force the small singular values to increase, and the increment is
proportional to $\sqrt{m}$. In our case where $\E$ is a sparse matrix
itself, we conjecture that the increment must be $O(1)$.
\end{remark}

$\tilde{\A}$ corresponds to the same bipartite graph as $\A$.
 However, the edges are labeled with positive fractional weights
 between 0 and 1, rather than single 1 weights of $\A$ . Besides, all
 the edges emanating from any node in $X$ have a weight sum-up equal
 to $d$. It is worth noticing that, after modifying $\A$ based on
 perturbations described above,  Theorem
\ref{Null Space}, Lemma \ref{lem:|w+|=|w-|} and Theorems \ref{thm:null
space for regular graphs} and \ref{thm:positive negative ratio} all
continue to hold for this class of matrices $\tilde{\A}$. Therefore
 $\Cr(\tilde{\A})\geq r_0$ will guarantee perfect recovery of
 $\frac{r_0}{d}$-sparse vectors via $\ell_1$-minimization. Moreover,
 the fact that $\Cr(\tilde{\A})\geq r_0$ can be translated back as if
 $\A$ is a ($r_0$,$\frac{d-1}{d}$) unbalanced expander
 graph. Therefore what we really care about is constructing
 ($r_0$,$\frac{d-1}{d}$)
expanders with $\frac{r_0}{nd}$ as large as possible. In section
 \ref{Sec:pobabilistic approach}, we use a probabilistic method to
 show that the desired ($r_0=\mu n$,$\frac{d-1}{d}$) expanders exist
 and give thresholds on $\frac{\mu}{d}$. Before continuing, note
 that we are using a $1-\epsilon = \frac{1}{d}$ expansion coefficient
 for perfect recovery, which is very small compared to other schemes that use
 expanders (see, e.g., \cite{Weiyu expander algorithm,Indyk
 RIP,Indyk_Ruzic,Indyk nullspace,XH07b,Weiyu_Exp2}) and require
 expansion coefficients at least larger than $1-\epsilon\geq
 \frac{3}{4}$. $1-\epsilon = \frac{1}{d}$ is indeed the \emph{critical}
 expansion coefficient. We shortly digress in a subsection to discuss
 this a little further.

\subsection{Necessity of Expansion} \label{sec: minimal expansion}

Consider any ${m\times n}$ measurement matrix $A$ that allows the recovery of all $r_0$-sparse vectors and construct its corresponding bipartite graph $B(n,m)$ by placing an edge between nodes $i,j$ if the $A_{ij}$ entry is nonzero.
We show that any such matrix must correspond to an expander bipartite graph.
The intuition is that a contracting set is certificate for a submatrix being rank-deficient, and hence reconstruction is impossible.
In particular, for the case where each column of $A$ has $d$ non-zero entries we obtain the following statement:
\begin{lem}
Any ${m\times n}$ measurement matrix $\A$ with $d$ non-zeros per column that allows the recovery of all $r_0$-sparse vectors must correspond to a ($r_0$,$1-\frac{1}{d}$) bipartite expander graph.
\end{lem}
\begin{proof}
The statement holds for any recovery algorithm (even $\ell_0$ minimization), because
we show that even a stronger recovery algorithm that is given the support of the $r_0$ sparse vector will fail to recover. Assume that the bipartite graph is not
a ($r_0$,$1-\frac{1}{d}$) expander, i.e. there exists a set of $r\leq r_0$
columns that is adjacent to $r-1$ (or less) rows. Therefore the submatrix corresponding to these $r$ columns must have rank strictly smaller than $r$ regardless of what the non-zero entries are.
By selecting a sparse signal that is supported exactly on these $r$ columns we see that it is impossible for any algorithm to recover it, even if the support is known, since there is rank-loss in the corresponding measurement submatrix.

This argument is easily extended to the non-regular case where the number of non-zeros in every column is arbitrary. The key necessary condition is that every set of size $r\leq r_0$ has a neighborhood of size $r$ or greater. This is exactly the condition of Hall's marriage theorem that guarantees a perfect matching for every subset of size up to $r_0$. This perfect matching combined with perturbations will
suffice to ensure that all the submatrices are full rank.
\end{proof}

%

\noindent Gaussian and other dense measurement matrices correspond to completely connected bipartite graphs which obviously have the necessary expansion. Therefore, the previous lemma becomes interesting when one tries to construct sparse measurement matrices.

We can easily show that bigger expansion is also sufficient (but not necessary); given the adjacency matrix of a general ($k$,$\epsilon$)-expander, can we
guarantee that with appropriate perturbations, $\ell_1$-optimization
will allow the recovery of $f(k,\epsilon)$-sparse vectors for some
positive function $f(.,.)$. The answer is yes, and the proof leverages
on the fact that every ($k$,$\epsilon$) graph can be interpreted as a
($k'$,$1-\frac{1}{d}$)-expander for some $k' \geq k$.

\begin{lem}\label{lem: multiple expansion}
If $\epsilon \leq 1-\frac{1}{d}$ and $k>0$, then every $d$-left regular ($k$,$\epsilon$)-expander is also a ($k(1-\epsilon)d$,$1-\frac{1}{d}$)-expander
\end{lem}
\begin{proof}
Consider a subset $S\subset X$ of the nodes with $|S|\leq k(1-\epsilon)d$. If $|S|\leq k $ then from the ($k$,$\epsilon$)-expansion $|\Gamma(S)|\geq (1-\epsilon)d|S|\geq |S|$. If $k<|S|\leq k(1-\epsilon)d$, then by choosing an arbitrary subset $S'\subset S$ with $|S'|=k$ and using the ($k$,$\epsilon$)-expansion on $S'$ we have:
\begin{eqnarray*}
|\Gamma(S)|\geq |\Gamma(S')|\geq k(1-\epsilon)d\geq|S|.
\end{eqnarray*}
\end{proof}
\vspace*{-5pt}
\noindent
Recall that a ($r_0$,$1-\frac{1}{d}$)-expander, when added with perturbations was capable of recovering $r_0$-sparse vectors. This and Lemma \ref{lem: multiple expansion} immediately imply:
\begin{thm}\label{thm: general expanders performance}
If $\A$ is the adjacency matrix of an unbalanced $d$-left regular
($k$,$\epsilon$)-expander graph, then there exists a perturbation of
$\A$ in the nonzero entries resulting in a nonnegative
matrix $\tilde{\A}$, such that every nonnegative
$k(1-\epsilon)$-sparse vector $\x$ can be recovered from
$\y=\tilde{\A}\x$ without error using $\ell_1$-optimization.
\end{thm}

This is an improvement over the existing bounds. For example,
\cite{Indyk nullspace}, guarantees sparse $\frac{k}{2}$ vectors can be
reconstructed via $\ell_1$-optimization using ($k$,$\epsilon$)
expanders with $\epsilon \leq \frac{1}{6}$. Using the above theorem
with $\epsilon=\frac{1}{6}$, $\frac{5}{6}k$-sparse non-negative
vectors can be perfectly recovered by linear programming. Likewise
\cite{Weiyu expander algorithm} provides an algorithm that recovers
vectors with sparsity $\frac{k}{2}$ using ($k$,$\epsilon$) expander
graphs with $\epsilon\leq\frac{1}{4}$. Our theorem for $\epsilon =
\frac{1}{4}$ allows for the recovery of $\frac{3}{4}k$-sparse
positive signals. Note that these bounds are very conservative and in
fact, the size of expanding sets when smaller expansion is required
are much bigger, which yields much bigger provably recoverable sets.

\section{Existence of Sparse Matrices with Linear Complete Rank}\label{Sec:pobabilistic approach}

For fixed values of $n>m>r_0$ and $d$ we
are interested in the question of the existence of
($r_0$,$\epsilon=\frac{d-1}{d}$)
expanders with constant left degree d. We use the standard first moment method argument  to prove the existence of such an expander
for appropriate ratios of $n$,$m$ and $r_0$ and $d$. The main result is given below, while the complete proof can be found in Appendix \ref{app:prob}.

\begin{thm}\label{thm: d mu beta}
    For sufficiently large $n$, with $m=\beta n$ and $r_0=\mu n$, there exists a bipartite graph with left vertex size $n$ and right size $m$ which is a ($r_0,\frac{d-1}{d}$) expander, if
\begin{equation}\label{eq:basic equation}
d>\frac{H(\mu)+\beta H(\frac{\mu}{\beta})}{\mu\log(\frac{\beta}{\mu})}.
\end{equation}
\end{thm}

\begin{figure}
  \centering
  \subfloat[Strong bound.]{\label{fig: strong bound}\includegraphics[width=0.4\textwidth]{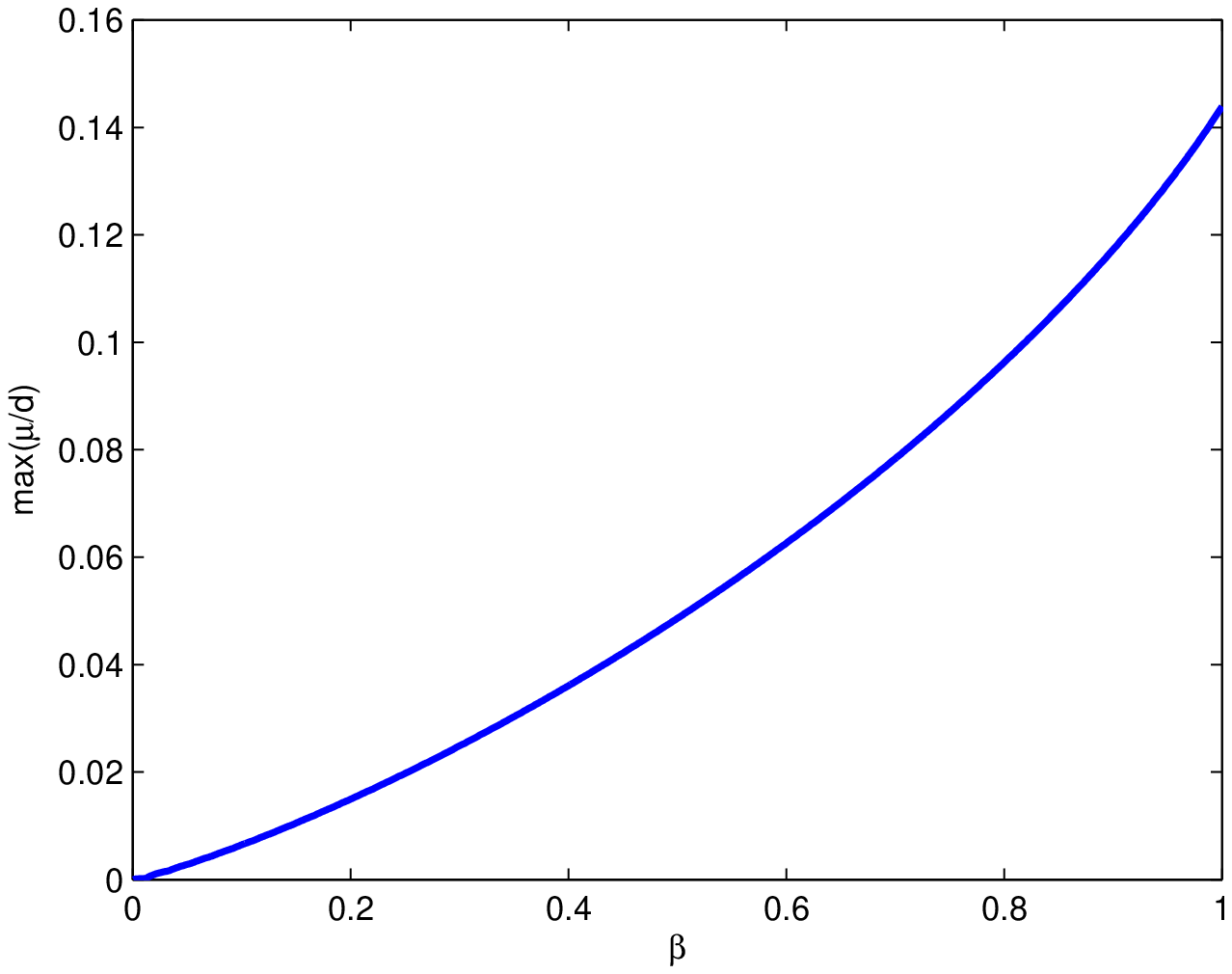}}
  \subfloat[Weak vs strong bound.]{\label{fig: weak vs strong}\includegraphics[width=0.4\textwidth]{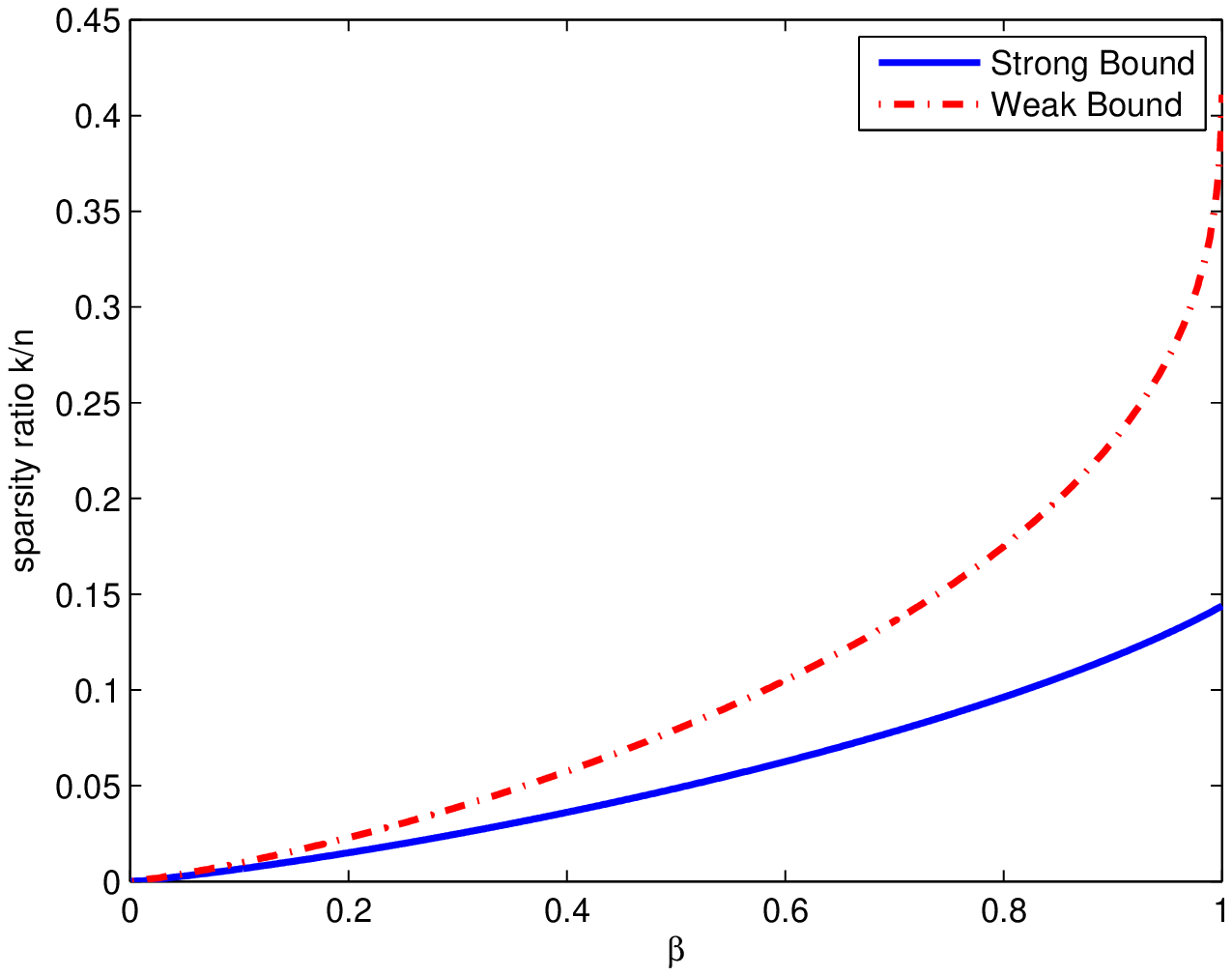}}
    \caption{ \small Comparison of weak achievable bound of Section \ref{Sec: weak bound} and with the strong achievable threshold of (\ref{eq:basic equation}) for $\frac{\mu}{d}$.}
  \label{fig: weak bound vs strong bound}
\end{figure}

More important is the
question of how big the ratio $\frac{\mu}{d}$ can be, since we
earlier proved that we can recover up to $\frac{r_0}{d}=\frac{\mu}{d}n$ sparse
vectors using this method. Figure \ref{fig: strong bound} illustrates the
maximum achievable ratio for different values of $\beta$  derived from (\ref{eq:basic equation}).

%
%

\subsection{Weak bound}
\label{Sec: weak bound}

We are now interested in deriving conditions for recovering a specific support set $S$ of size $k=\alpha n$, rather than obtain a worst case bound for matrices that work for all support sets.
Recall that $m= \beta n$, left degree is $d$, and define $\gamma_1:=(1-e^{-d\frac{\alpha}{\beta}}) \beta$.
\begin{thm}\label{thm:weakbound}
Define the function
\beq
F(\rho_1,\rho_2):= \alpha H(\frac{\rho_1}{\alpha})+ (1-\alpha) H(\frac{\rho_2}{1-\alpha})+\beta H(\frac{\rho_1+\rho_2}{\beta})+d(\rho_1+\rho_2) \log(\frac{\rho_1+\rho_2}{\beta}).
\eeq
For every $\alpha$ such that $F(\rho_1,\rho_2)<0$ for every
$\rho_1,\rho_2$ that satisfies $\rho_1 <\alpha, \rho_2<1-\alpha,
\rho_1+\rho_2 <\gamma_1$, a randomly selected subset of size $k=\alpha
n$ is recoverable from a random perturbed matrix $\tilde{\A}$ with
probability $1-o(1)$.
\end{thm}
The bound that results from Theorem \ref{thm:weakbound} is plotted in Figure \ref{fig: weak vs strong} and has been compared to the strong threshold previously achieved. The full proof of this statement is given in Appendix \ref{app:weak}.
The key argument is a matching condition for the recoverability of vectors supported on a specific subset $S$. The condition involves looking at the two-hop graph from $S$ and checking if all sets of size up to $|\Gamma(S)|+1$ have a perfect matching:
\begin{lem}
Given a set $S$ consider $\Gamma(S)$ and denote $S_2= \Gamma(\Gamma(S)) \setminus S$. Let the bipartite two-hop graph of $S$ be denoted by $B_{S}=(S \cup S_2, \Gamma(S \cup S_2))$. Any non-negative vector $\x_0$ supported on $S$ can be recovered from $\y=\tilde{\A}\x_0$, if every subset $S' \subset S \cup S_2$ of size $|S'|\leq |\Gamma(S)|+1$ has minimal expansion: $|\Gamma(S')|\geq |S'|$.
\end{lem}
\begin{proof} 
Consider the two-hop bipartite graph of $S$ and let $C=(S \cup S_2)^c$ denote the remainder of the nodes in $X$. Further let $\A_S$ denote the  submatrix corresponding to $B_S$.
 By Hall's theorem since every subset of $B_{S}$ of size up to $|\Gamma(S)|+1$ has expansion equal to its size, it must also have a perfect matching and by the same perturbation argument $\Cr(\tilde{A_S})\geq |\Gamma(S)|$.

By our null space characterization, to show that a set $S$ can be recovered, it suffices to show that every nonzero vector $\w$ in the nullspace of $\tilde{A}$ cannot have all its negative components in $S$.
Assume otherwise: that some $\w$ has indeed all its negative support $S_{\w}^- \subseteq S$. Observe now that $C$ cannot contain any of the positive support of $\w$, because every equation that is adjacent to a positive element must also be adjacent to a negative (since the matrix coefficients are positive) and $\Gamma(S_{\w}^-)$ does not intersect $\Gamma(C)$. Therefore the whole support of $\w$ must be contained in $S\cup S_2$.

Now we can show that $|S_{\w}| \leq |\Gamma(S)|$.
Assume otherwise, that $|S_{\w}| > |\Gamma(S)|$. Then we could select a subset of $K \subseteq S_{\w}$ such that $|K|=|\Gamma(S)|+1$. This set $K$ satisfies our assumption and is contained in $B_S$ and therefore must have the minimal expansion $|\Gamma(K)|\geq |K|=|\Gamma(S)|+1$. But since
$\Gamma(S_{\w}^-)=\Gamma(S_{\w}^+)=\Gamma(S_{\w})$ and  $K \subseteq S_{\w}\subseteq S$,
it must hold that $|\Gamma(K)|\leq |\Gamma(S)|$, which contradicts the minimal expansion inequality.

Therefore, $|S_{\w}|$ must have a perfect matching which means that we can find a full rank submatrix $\A_{\w}$ (corresponding to that matching) such that $\A_{\w} \w_S =0$ (where by $\w_S$ we denote the vector $\w$ restricted to its support). Since $\A_{\w}$ is full rank, $\w$ must be the all zeros vector which contradicts the assumption that $S_{\w}^{-}$ can be contained in $S$.
\end{proof}


\section{Fast Algorithm} \label{sec: fast alg.}
We now describe a fast algorithm for the
recovery of sparse non-negative vectors from noiseless
measurements. This algorithm relies on the minimal expansion  we
described in section \ref{sec: minimal expansion}. We
employ a ($kd+1$,$1-\frac{1}{d}$) expander and perturb it as Lemma
 (\ref{lem: perturbations}) to obtain a sparse nonnegative matrix
$\tilde{\A}$ with $\Cr(\tilde{\A})\geq kd+1$. Knowing that the target signal is at most $k$-sparse the algorithm works as follows

\begin{alg}{Reverse Expansion Recovery}\label{algo: alg1}
\begin{enumerate}

\item Find zero entries of $\y$ and denote them by $\y_1$. Also denote
      by $T_1$ the index set of elements of $\y_1$ in $\y$, and by
      $T_2$ its complement. Wlog assume that
      $\y=\left[\begin{array}{c}
      \y_1 \\ \y_2 \end{array}\right]$.
\item Locate in $X$ the neighbors of the set of nodes in $Y$ corresponding to $T_1$, name the set $S_1$ and name the set of their complement nodes in $X$ by $S_2$.

\item Identify the sub-matrix of $\tilde{\A}$ that represents the nodes emanating from $S_2$ to $T_2$. Call this sub-matrix $\tilde{\A_2}$. Columns of $\tilde{\A_2}$ correspond to nodes in $S_2$, and its rows correspond to the nodes in $T_2$.

\item Set $\hat{\x}_{S_1}=0$ and compute $\hat{\x}_{S_2}={\tilde{\A}_2}^\dag \y_2$, where $\A^\dag$ is the pseudo-inverse of $\A$ defined by $\A^\dag=(\A^t \A)^{-1}\A^t$.  Declare $\hat{\x}$ as the output.

\end{enumerate}

\end{alg}

The algorithm begins with identifying a big zero portion of the
output and locating their corresponding nodes in $Y$ (Refer to Figure
\ref{fig:bipgraph} in Appendix \ref{app:alg}). In the next
step, neighbors of these nodes are found in $X$ and these two giant
sets of nodes are eliminated from $X$ and $Y$. Having done this, we
are left with a much smaller system of linear equations, which turns
out to be \emph{over-determined}, and therefore our problem reduces to solving linear equations.
The theoretical justification of the above
statement, and why the algorithm works, is provided in Appendix
\ref{app:alg}.
\subsection{Noisy Case, Robust Algorithm }
In general, the observations vector is contaminated with measurement
noise, usually of significantly smaller power, leading to the
following equation:
\vspace*{-5pt}
\begin{equation}\label{eq: noisy obs.}
\y=\A\x+\v.
\end{equation}
As before, it is assumed in (\ref{eq: noisy obs.}) that $\x$ is
sparse. $\v$ is the $n\times 1$ observation noise vector with limited
$\ell_1$-norm. In practice $\v$ is often characterized by its $\ell_2$
norm. However, in order to establish a recovery scheme that is robust
to a limited power noise, we need to have a measurement matrix with a
$2$-RIP. This is not in general true for $(0,1)$ expander graphs,
although it is realizable via a suitable perturbation for
this class of matrices. However, for the scope of this paper, we
assume that the limitation on the noise is given through its
$\ell_1$-norm. This allows the of use the $1$-RIP bounds of
\cite{Indyk RIP} to analyze the performance of our scheme in the
presence of noise. Again we are assuming that $\x$ is $k$-sparse\\
\begin{alg}{Noisy Case}\label{algo: noisy case}
\begin{enumerate}
\item Sort elements of $\y$ in terms of absolute value, pick the smallest $m-kd$ ones and stack them in a vector denoted by $\y_1$. Also denote by $T_1$ the index set of elements of $\y_1$ in $\y$, and by $T_2$ its complement. Wlog assume that $\y=\left[\begin{array}{c}
      \y_1 \\ \y_2 \end{array}\right]$.
      \vspace*{-5pt}
\item Locate in $X$ the neighbors of the set of nodes in $Y$ corresponding to $T_1$, name them by $S_1$ and name the set of their complement nodes in $X$ by $S_2$.
\vspace*{-5pt}
\item Identify the sub-matrix of $\A$ that represents the nodes emanating from $S_2$ to $T_2$ referred to as $\A_2$. Columns of $\A_2$ correspond to nodes in $S_2$, and its rows stand for the nodes in $T_2$.
\vspace*{-5pt}
\item Set $\hat{\x}_{S_1}=0$ and $\hat{\x}_{S_2}= \mbox{argmin}_{\z\in \mathbb{R}^{|S_2|\times 1}
}{\|\A_2 \z-\y_2\|_1}$ and declare $\hat{\x}$ as the output.

\end{enumerate}
\end{alg}

We show in the following theorem that our algorithm is robust (in a $\ell_1$-norm sense) to the observation noise.

\begin{thm}\label{thm: robustness of alg}
If $\A$ is the adjacency matrix of a ($k$,$\epsilon$)-expander with $\epsilon < 0.5$, $\x$ is a $k$-sparse nonnegative vector and $\hat{\x}$ is the output of
Algorithm \ref{algo: noisy case}, then $\|\x-\hat{\x}\|_1 \leq \frac{6-4\epsilon}{1-2\epsilon}\|\v\|_1$
\end{thm}
\vspace{-5pt}
\begin{proof}
Given in Appendix \ref{app:noise}.
\end{proof}

%

\vspace*{-5pt}

\section{Experimental Evaluation}
\begin{figure}
  \centering
  \subfloat[\scriptsize $\ell_1$-minimization on expanders and perturbed expanders.]{\label{fig:recovery percentage}\includegraphics[width=0.4\textwidth]{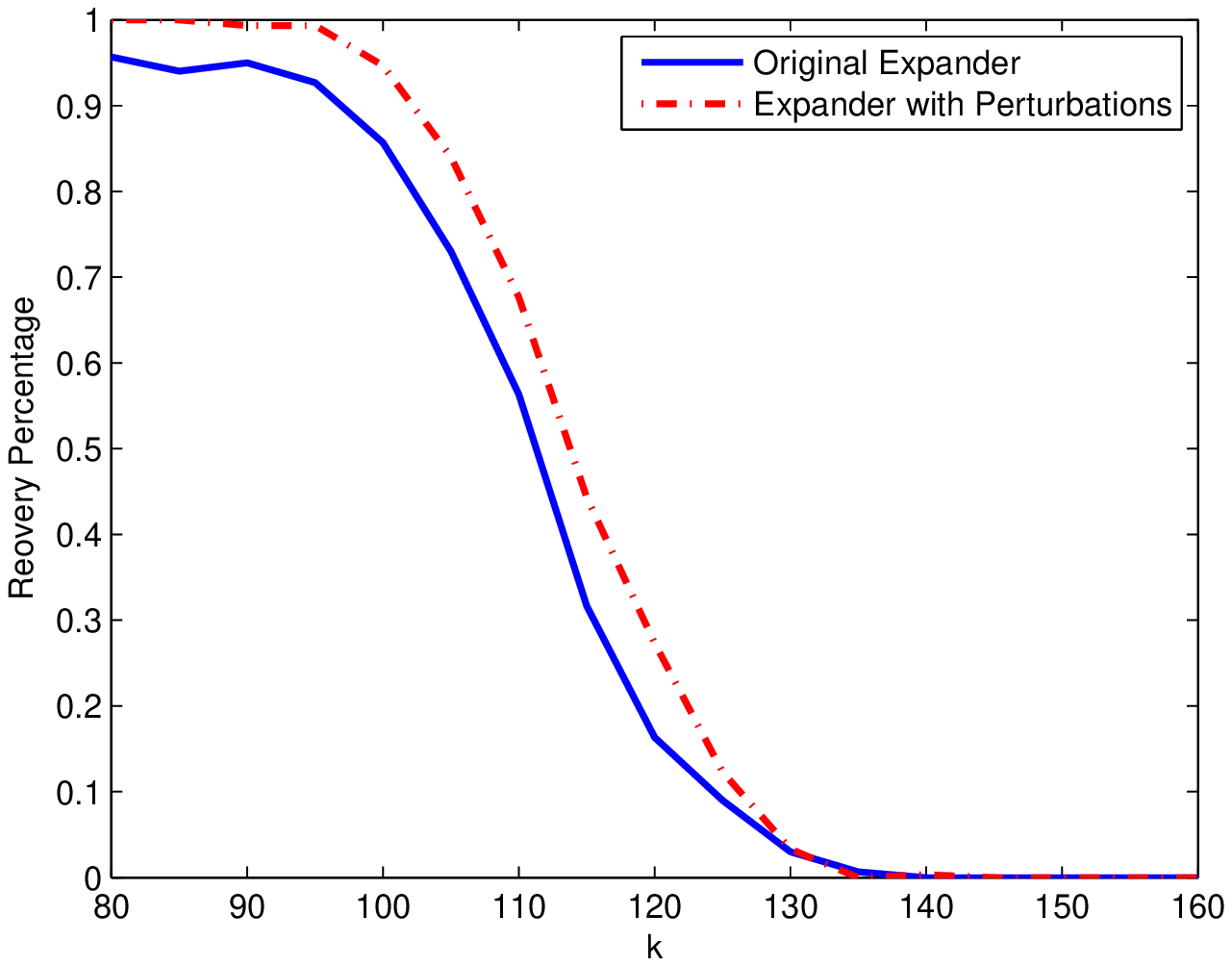}}
  \subfloat[\scriptsize Algorithm \ref{algo: alg1} Vs. $\ell_1$-minimization.]{\label{fig: performance of our algorithm}\includegraphics[width=0.4\textwidth]{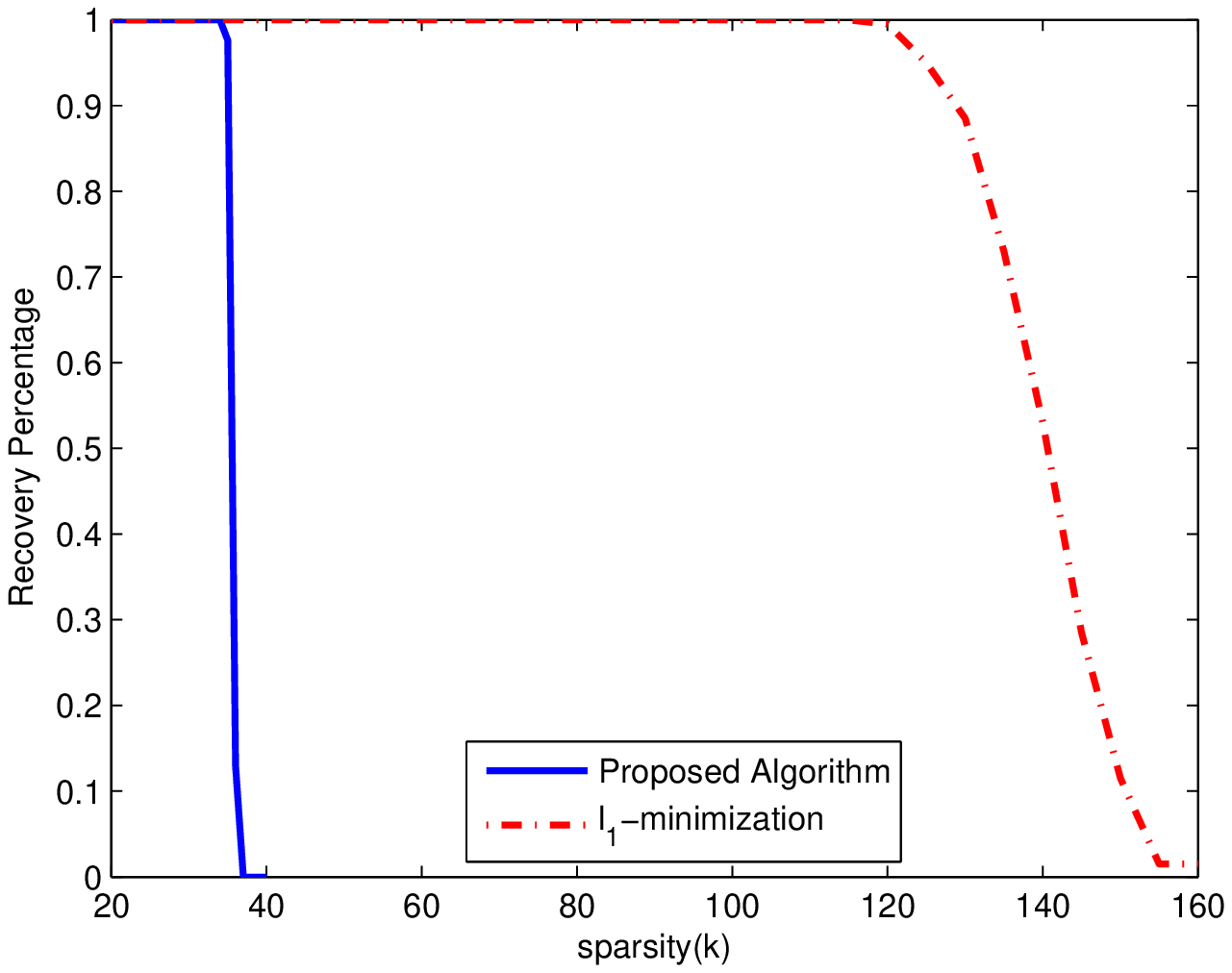}}
    \caption{ \scriptsize Percentage of successful recovery for $\ell_1$-minimization and Algorithm \ref{algo: alg1} with minimal expander measurements and reconstruction error of Algorithm \ref{algo: noisy case}.}
  \label{fig: comparison of recovery erformances}
\end{figure}

We generated a random $m\times n$ matrix $\A$ with $n=2m=500$, and
$d=3$ 1's in each column. We then multiplied random sparse vectors
with different sparsity levels by $\A$, and tried recovering them via
the linear programming (\ref{LP}). Next, we added the perturbations
described in section \ref{Sec: Completer Rank} to $\A$ and applied the
same sparse vectors to compare the recovery percentages in the two
cases. This process was repeated for a few generations of $\A$ and the
best of the improvements we obtained is illustrated in
Figure \ref{fig:recovery percentage}.

In Figure \ref{fig: performance of our algorithm}  we plot the
recovery percentage of Algorithm \ref{algo: alg1} for a random
perturbed expander adjacency matrix $\tilde{A}$ of size $250\times
500$ and $d=6$, and we have compared the performance with the
$\ell_1$-minimization method. Although the deterministic theoretical bounds of the two methods are
the same, as observed in Figure \ref{fig: performance of our
  algorithm}, in practice $\ell_1$-minimization is more effective for
less sparse signals. However Algorithm \ref{algo: alg1} is considerably
faster than linear programming and easier to implement.

\begin{figure}
\centering
  \includegraphics[width=0.7\textwidth]{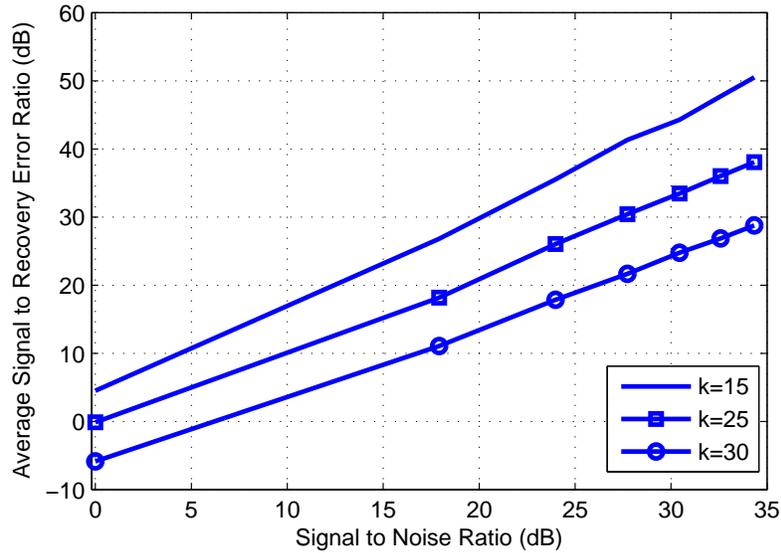}\\
\caption{Simulation results for Algorithm \ref{algo: noisy case}; Signal to Error Ratio Vs. SNR.}
\label{fig: SER vs. SNR}
\end{figure}

In general, the complexity
of Algorithm \ref{algo: alg1} is $O(nk^2)$
which, when $k$ is proportional to $n$, is similar to linear
programming $O(n^3)$. However the constants are much smaller,
which is of practical advantage. Furthermore, taking advantage
of fast matrix inversion algorithms for very sparse matrices,
Algorithm \ref{algo: alg1} can be performed in dramatically less
operations. Figure \ref{fig: SER vs. SNR} shows the Signal to Error
Ratio as a function of Signal to Noise Ratio when Algorithm \ref{algo:
  noisy case} (with $\ell_2$-norm used in step $4$) has been used to
recover noisy observations. Assuming that the output of the algorithm is $\hat{\x}$, Signal to Noise Ratio (SNR) and Signal to Error Ratio functions are defined as
\begin{eqnarray}
SNR = 10\log\frac{\|\A\x\|_2^2}{\|\v\|_2^2}   \\
SER = 10\log\frac{\|\x\|_2^2}{\|\x-\hat{\x}\|_2^2}
\end{eqnarray}
Measurement matrices are the same as
before.

\section{Conclusion}

In this paper we considered the recovery of a non-negative sparse
vector using a sparse measurement matrix in the compressed sensing
framework. We used the perturbed adjacency matrix of a bipartite
expander graph to construct the sparse measurement matrix and proposed
a novel fast algorithm. We computed recovery thresholds and showed
that for measurement matrices with non-negative entries and constant
column sum the constraint set $\{\x|\x\geq 0,\A\x=\y\}$ is unique
whenever $\ell_1$ optimization is successful (which also means that
any other optimization scheme would be successful). Finally,
determining whether the matrices constructed satisfy an 2-RIP (we
conjecture that they do), and constructing $(0,1)$ matrices that have
complete rank proportional to $n$ are open problems that may be worthy
of further scrutiny.

\renewcommand{\theequation}{\thesection.\arabic{equation}}

\appendix{}
\renewcommand{\theequation}{\thesection.\arabic{equation}}

\renewcommand{\theequation}{\thesection.\arabic{equation}}
\section{Proof of Theorem \ref{thm: d mu beta}}
\label{app:prob}

 Assuming that we generate a random matrix $\A$ by
randomly generating its columns, it suffices to show that the
probability that $\A$ has the desired expansion property is
positive. For $1\leq i_1< i_2 <...< i_r\leq n$ We denote by
$E_{i_1,i_2,...,i_r}$ the event that the columns of $\A$
corresponding to the numbers $i_1$,$i_2$,...,$i_r$ have at
least $n-r-1$ entire 0 rows (rows that does not have a single
non-zero elements in the columns $\A_{i_1}$,
$\A_{i_2}$,...$\A_{i_k})$. In other words $E_{i_1,i_2,...,i_r}$ is the event that the set of nodes $\{i_1,i_2,...,i_r\}$ in $X$ contracts in $Y$.

\vspace{2pt}
\begin{eqnarray}
\nonumber \Prob[ \A \:\: \text{is a ($r_0$, $\frac{d-1}{d}$)-Exp.}] &=&  1-\Prob[\A \:\: \text{not a ($r_0$,$\frac{d-1}{d}$-Exp.)} ]\\
\nonumber &=& 1 - \Prob[\bigcup_{d\leq r\leq r_0,1\leq
i_1<i_2<...<i_r}{E_{i_1,i_2,...,i_r}}]\\
\nonumber &=& 1 - \sum_{r=d}^{r_0}{n\choose r}{\Prob[E_{1,2,...,r}]}
\end{eqnarray}
A combinatorial analysis yields the following:
\begin{eqnarray}
\nonumber \Prob[E_{1,2,...,r}]\leq\frac{{m\choose r}{r\choose d}^r}{{m\choose d}^r}
\end{eqnarray}
Hence
\begin{eqnarray}\label{eq:Upper bound}
\Prob[A \:\: \text{is a ($r_0$,$\frac{d-1}{d}$)-Exp.}]\geq
1-\sum_{r=d}^{r_0}{{n\choose r}\frac{{m\choose r}{r\choose
d}^r}{{m\choose d}^r}}
\end{eqnarray}

The objective is to show that this probability is
positive. Equivalently, we show that for certain regimes of $\beta$,
$\mu$ and $d$, the summation on R.H.S of (\ref{eq:Upper bound})
vanishes. To this end, we split the sum into a sub-linear summation
and a summation. We show that if $d>2$, the sub-linear part will
decays polynomially as $n\rightarrow \infty$, and the linear part
decays exponentially in $n$, provided a certain relationship involving
$\beta$,$\mu$ and $d$ holds. We state this in two different theorems.

\begin{thm}\label{thm: sub-linear}
If $0<\alpha <e^{\frac{d}{2-d}}\beta^{\frac{1-d}{2-d}}$ and $d\geq3$ then $\sum_{r=d}^{\alpha n}{{n\choose r}\frac{{m\choose r}{r\choose
d}^r}{{m\choose d}^r}}=O(n^{1-d(d-2)})$.
\end{thm}
\begin{proof}
We can write:

\begin{eqnarray}
\sum_{r=d}^{\alpha n}{{n\choose r}\frac{{m\choose r}{r\choose d}^r}{{m\choose d}^r}} &\leq&
\sum_{r=d}^{\alpha n}{{n\choose r}{m\choose r}(\frac{r}{m})^{rd}}  \\ \label{eq:sub-linear 1}
&\leq& \sum_{r=d}^{\alpha n}{(\frac{ne}{r})^r(\frac{me}{r})^r(\frac{r}{m})^{rd}} \\ \label{eq:sub-linear 2}
&=& \sum_{r=d}^{\alpha n}{(\frac{cr}{n})^{r(d-2)}} \label{eq:sub-linear 3}
\end{eqnarray}
Where $c=e^{\frac{2}{d-2}}\beta^{\frac{1-d}{d-2}}$. (\ref{eq:sub-linear 1}) and (\ref{eq:sub-linear 2}) are deduced from the bounds $\frac{{r\choose d}}{{m\choose d}} \leq (\frac{r}{m})^d$ for $r<m$, and ${n\choose k}\leq (\frac{ne}{k})^k$ respectively. It is easy to show that when $\alpha < \frac{1}{ec}$, $(\frac{cr}{n})^r$ is decreasing in $r$, and thus replacing all the terms in (\ref{eq:sub-linear 3}) by the first term will only increase the sum. The whole term is thus smaller than $\alpha n (\frac{cd}{n})^{d(d-2)}=\lambda n^{1-d(d-2)}$ for some positive constant $\lambda$.
\end{proof}

\begin{thm}\label{thm: linear}
For $m=\beta n$ and  $r_0=\mu n$, if  $d>\frac{H(\mu)+\beta H(\frac{\mu}{\beta})}{\mu\log(\frac{\beta}{\mu})}$ then for any $0<\alpha<\mu$, the sum $\sum_{r=\alpha n+1}^{\mu n}{{n\choose r}\frac{{m\choose r}{r\choose d}^r}{{m\choose d}^r}}$ decays exponentially as $n\longrightarrow \infty$.
\end{thm}
\begin{proof}
Using the standard bounds of (\ref{eq: binomial bounds}) on binomial coefficients we can write:
\begin{equation}\label{eq: linear bound}
\sum_{r=\alpha n+1}^{n}{{n\choose r}\frac{{m\choose r}{r\choose d}^r}{{m\choose d}^r}}\leq n^2\sum_{r=\alpha n+1}^{\mu n}{2^{nH(\frac{r}{n})+{mH(\frac{r}{m})}+r^2H(\frac{d}{r})-mrH(\frac{d}{m})}}
\end{equation}
where $H(x)=x\log(\frac{1}{x})+(1-x)\log(\frac{1}{1-x})$ is the entropy function.
Using  $H(\epsilon)=
\epsilon(\log(\frac{1}{\epsilon})+1)+O(\epsilon^2)$ for small $\epsilon$ and the fact that as $n\rightarrow\infty$ $\frac{d}{r}\rightarrow 0 $ and $\frac{d}{m}\rightarrow 0$ for $\alpha n< r\leq \mu n$, (\ref{eq: binomial bounds}) can be written as:
\begin{eqnarray}
\nonumber \sum_{r=\alpha n+1}^{n}{{n\choose r}\frac{{m\choose r}{r\choose d}^r}{{m\choose d}^r}} &\leq& n^2\sum_{r=\alpha n+1}^{\mu n}{2^{nH(\frac{r}{n})+{mH(\frac{r}{m})}+rd\log_2{\frac{d}{m}}+O(1)}} \\
&=& O(n^3 2^{n(H(\mu)+\beta H(\frac{\mu}{\beta})+\mu d\log(\frac{\mu}{\beta}))}) \label{eq: exponent}
\end{eqnarray}
(\ref{eq: exponent}) vanishes if $d>\frac{H(\mu)+\beta H(\frac{\mu}{\beta})}{\mu\log(\frac{\beta}{\mu})}$.
\end{proof}
The above argument leads to Theorem \ref{thm: d mu beta}.

\renewcommand{\theequation}{\thesection.\arabic{equation}}
\section{Derivation of the Weak Bound}
\label{app:weak}
We start with a straightforward modification of Theorem \ref{Null Space}:
\begin{thm}\label{thm:weak bound}
let $\x_0\in (\mathbb{R}^+)^n$ be fixed, $\y=\A\x_0$ and denote by $S$ the support of $\x_0$. Then the solution $\x$ of (\ref{LP}) will be identical to $\x_0$ if and only if there exists no $\w$ in the null space of $\A$ such that $\w_{S^c}$ is nonnegative and $\sum_{i=1}^{n}w_i > 0$.
\end{thm}
\begin{proof}
Similar to the proof of  Theorem \ref{Null Space} with $S$ as the support of $\x_0$
\end{proof}

\begin{cor}\label{cor:weak bound}
If $\A$ is the adjacency matrix of a bipartite graph with left constant degree, and if $\x_0$ is a fixed nonnegative vector and $\y=\A\x_0$,  then the solution $\x$ of (\ref{LP}) will be identical to $\x_0$ if and only if there exists no $\w$ in the null space of $\A$ so that $\w_S^c$ is a nonnegative vector, where $S$ is the support set of $\x_0$. In other words $\x_0$ will be recoverable via L.P from $\A\x_0$ provided the support of $\x_0$ does not include the index set of all negative elements of a vector in the null-space of $\A$.
\end{cor}
\begin{proof}
Directly from Theorem \ref{thm:weak bound} and Lemma \ref{lem:|w+|=|w-|}.
\end{proof}

This last statement allows us to derive a combinatorial matching condition for the recovery of a vector supported on a specific subset $S$. We repeat the statement of the lemma:
\begin{lem}
Given a set $S$ consider $\Gamma(S)$ and denote $S_2= \Gamma(\Gamma(S)) \setminus S$. Let the bipartite two-hop graph of $S$ be denoted by $B_{S}=(S \cup S_2, \Gamma(S \cup S_2))$. Any non-negative vector $\x_0$ supported on $S$ can be recovered from $\y=\tilde{\A}\x_0$, if every subset $S' \subset S \cup S_2$ of size $|S'|\leq |\Gamma(S)|+1$ has minimal expansion: $|\Gamma(S')|\geq |S'|$.
\end{lem}
%
%
%

Observe that the expectation is (asymptotically)
$\stexp \Gamma(S)= (1-e^{-d\frac{|S|}{m}}) \beta n =: \gamma_1 n$
Using a standard Chernoff bound ~\cite{Chernoff bound}  it is easy to show that $\Gamma(S)$ is concentrated around its expectation:
\[ \Prob[ \Gamma(S) <=\stexp \Gamma(S)+ \epsilon_1] > 1-\frac{1}{n},\]
if $|S|\geq c_1 n$, for $\epsilon_1= c_2 \sqrt{n \log n}$.
Therefore we define the event $E_1= \{ \Gamma(S) <=\gamma_1 n+ \epsilon_1\}$.

Consider the random graph created from placing $d$ non-zero entries (with repetition) in every column of $\tilde{A}$.
From the set $S$, form $\Gamma(S)$, the corresponding $S_{2}$, and finally the bipartite graph $B_{S}= (S \cup S_2 , \Gamma(S \cup S_2))$.
Using the given combinatorial condition, we can recover a signal supported on $S$ if every subset $S_i\subset S \cup S_2$ of size $|r| \leq |\Gamma(S)|+1$ has sufficient expansion: $|\Gamma(S_i)| \geq r$ (note that subsequently we drop the $+1$ term since it is negligible for large $n$).
First we condition on the concentration of $\Gamma(S)$:
\begin{align}
\Prob[ S \:\: \text{not recoverable}] &=  \Prob[ S  \:\: \text{not recoverable}| E_1] \Prob[E_1]+
\Prob[ S  \:\: \text{not recoverable}| E^c_1] \Prob[E^c_1] \\
&\leq  \Prob[ S  \:\: \text{not recoverable}| E_1] (1-\frac{1}{n}) + \frac{1}{n},
\end{align}
therefore it suffices to bound the probability conditioned on $\Gamma(S)$ concentrated.
We are going to do a union bound over all possible selections of $r_1$ nodes in $S$ and $r_2$ nodes in $S_2$ so that $r1+r2 \leq \Gamma(S)+\epsilon_1$. Since we are conditioning on $E_1$ it suffices to have
$r1+r2 \leq \stexp \gamma_1 n$. The second problem is that the set $S_2$ is random and dependent on $\Gamma(S)$. We are going to avoid this conditioning by allowing the choice of $r_2$ to range over all the nodes $n-k$ nodes in $S^c$.
\beq
\Prob[ S  \:\: \text{not recoverable}| E_1] \leq \sum_{r_1+r_2 \leq \gamma_1 n}
{k \choose r_1} {n-k \choose r_2} \Prob( r1,r2 \:\: \text{contracts}| E_1). \eeq

Now the problem is that conditioning on $E_1$ implies that the set $r_1$ does not expand too much, so  it is actually increasing the probability of the bad contraction event. We can however easily show that this increase is at most a factor of 2:
\begin{align}
\Prob( r1,r2 \:\: \text{contracts}| E_1)= \frac{ \Prob( r1,r2 \:\: \text{contracts} \cap E_1)}{\Prob(E_1)}  \leq \frac{\Prob( r1,r2 \:\: \text{contracts})}{\Prob(E_1)}.
\end{align}
Now since $\Prob(E_1)\geq 1-1/n$, for sufficiently large $n$, $1/ \Prob(E_1) \leq 2$, so
\begin{align}
\Prob( r1,r2 \:\: \text{contracts}| E_1) \leq 2 \Prob(r1,r2 \:\: \text{contracts}).
\end{align}
The probability that the set $r1,r2$ contracts can be further bounded by assuming $\Gamma(r1,r2)= r1+r2$ (any smaller neighborhood will have smaller probability) so
\[
\Prob(r1,r2 \:\: \text{contracts}) \leq {m \choose r_1+r_2}  \left(\frac{r_1+r_2}{m}\right)^{d(r_1+r_2)}.  \]
Putting everything together we obtain the bound
\beq
\Prob[ S  \:\: \text{not recoverable}| E_1] \leq 2 \sum_{r_1+r_2 \leq \gamma_1 n}
{k \choose r_1} {n-k \choose r_2} {m \choose r_1+r_2}  \left(\frac{r_1+r_2}{m}\right)^{d(r_1+r_2)}.
\eeq
We move everything to the exponent and use standard binomial approximations to obtain
\beq
\label{eq_probbound1}
\Prob[ S  \:\: \text{not recoverable}| E_1] \leq 2 \sum_{r_1+r_2 \leq \gamma_1 n}
2^{k H(\frac{r_1}{k})+ (n-k) H(\frac{r_2}{n-k})+m H(\frac{r_1+r_2}{m})+d(r_1+r_2) \log(\frac{r_1+r_2}{m})}. \eeq

Recall that the recoverable fraction is $k= \alpha n$, $m= \beta n$, and denote $\rho_1=r_1 /n$,
$\rho_2=r_2 /n$. Define the function
\beq
F(\rho_1,\rho_2):= \alpha H(\frac{\rho_1}{\alpha})+ (1-\alpha) H(\frac{\rho_2}{1-\alpha})+\beta H(\frac{\rho_1+\rho_2}{\beta})+d(\rho_1+\rho_2) \log(\frac{\rho_1+\rho_2}{\beta}),
\eeq
and observe that the bound on the probability of failure (\ref{eq_probbound1}) becomes
\[
\Prob[ S \:\: \text{not recoverable}| E_1] \leq 2 \sum_{r_1+r_2 \leq \gamma_1 n}
2^{n F(\rho_1,\rho_2)}.
\]
Therefore for a fixed $\beta, \gamma_1$ we are trying to find the largest $\alpha^*$ that makes
$F(\rho_1,\rho_2)$ negative for every $\rho_1,\rho_2$ for which $\rho_1+\rho_2\leq \gamma_1$.
For this $\alpha^*$, we can recover with exponentially high probability conditioned on the sublinear sets do not contract (which has been already established).

\renewcommand{\theequation}{\thesection.\arabic{equation}}
\section{Proof of the Validity of Algorithm \ref{algo: alg1}}
\label{app:alg}

\begin{figure}[h]
\centering
  \includegraphics[width=0.3\textwidth]{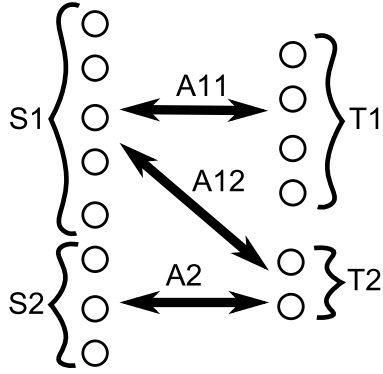}\\
\caption{Decomposition of nodes and edges by Algorithm \ref{algo: alg1}.}
\label{fig:bipgraph}
\end{figure}

As mentioned earlier and illustrated in Figure \ref{fig:bipgraph}, the algorithm begins by identifying a big zero
portion of the output and eliminating two large sets of nodes from $X$
and $Y$. Having done this, we
are left with a much smaller system of linear equation, which turns
out to be an over-determined system and can be uniquely solved using
matrix inversions. The theoretical justification of the above
statement, and why the algorithm works, is provided in Appendix
\ref{app:alg}.
The fact that it is over-determined is secured by
the expansion property of the measurement matrix and its proportional
complete rank.
Figure \ref{fig:bipgraph} is a graphical illustration of how Algorithm\ref{algo: alg1} decomposes $X$ and $Y$ into subsets and makes the search and observation spaces shrink from $X$ and $Y$ into $S_2$ and $T_2$ respectively. Intuitively speaking, when we isolate a big proportion of the nodes on the right, their neighbors in $X$ are big enough to leave us with a set of nodes on the left that are less than the remaining nodes on the right. This procedure is therefore nothing but a block diagonalization (by rearranging rows and columns)of $\tilde{\A}$
into a lower triangular matrix:
\begin{equation}\label{eq: diagonalization}
\left[\begin{array}{cc}
\tilde{\A}_{11} & 0 \\
\tilde{\A}_{12} & \tilde{\A_2}
\end{array}\right]
\end{equation}
where $\tilde{\A}_2$ is a square or tall full rank matrix. The following theorem certifies that Algorithm\ref{algo: alg1} is indeed valid and it recovers any $k$-sparse vector without error.

\begin{thm}{Validity of Algorithm}\label{thm: val of alg}

If $\x$ is a $k$-sparse non-negative vector and $\tilde{\A}$ is a perturbed ($kd+1$,$1-\frac{1}{d}$) expander with $\Cr(\A)\geq kd+1$ then:
\begin{enumerate}
\item  $\y$ is $kd$-sparse and therefore has at least $m-kd$ zeros.
\item $|S_2|\leq|T_2|$ and $\tilde{\A_2}$ is full rank.
\item $\hat{\x}=\x$
\end{enumerate}
\end{thm}

\begin{proof}{}
\indent
\begin{enumerate}
\item Trivial by noting the fact that the graph representation of $\tilde{\A}$ is left $d$-regular.
\item Suppose $|S_2|> |T_1|=kd$. select an arbitrary subset of ${S_2}'\subseteq S_2$ of size $kd+1$. Because of the expansion property:
 $|\Gamma({S_2}')|\geq kd+1>|T_1|$. But $\Gamma({S_2}')$ is in $T_2$ and this is a contradiction.  Diagonalization of  (\ref{eq: diagonalization}) and the fact that $\tilde{\A_2}$ is a tall matrix and $\Cr(\tilde{\A})\geq kd+1$ together imply that $\tilde{\A_2}$ has full column rank.

\item Every node in $Y$ is an equation with nonnegative variables from $\x$, and positive weights from edges of the graph. If any entry in $\x_{S_1}$ is greater than zero, then the equations corresponding to its neighbors in $T_1$ are not zero (and there is at least one such equation since $S_1 = \Gamma(T_1)$). This is in contradiction with the choice of $T_1$. Therefore $\x_{S_1}=0=\hat{\x}_{S_1}$. Also since $\tilde{\A_2}\x_{S_2}=\y_2$ and $\tilde{\A_2}\hat{\x}_{S_2}=\y_2$ and $\tilde{\A_2}$ is full rank we conclude that $\x_{S_2}=\hat{\x}_{S_2}$.

\end{enumerate}
\end{proof}

\begin{remark}
Note that in proving the last part of Theorem \ref{thm: val of alg}, we used the fact that $\x$ is a non-negative vector. This proof does not hold for general sparse vectors.
\end{remark}
\begin{remark}
By proving the part 2 of \ref{thm: val of alg} we implicitly proved that every expander from $X$ to $Y$ is a contractor from $Y$ to $X$. This can be generalized as following :
\begin{equation}
\forall T\subset Y \mbox{, } |T|\leq rd+1 \Rightarrow |\Gamma(T)|\leq |T|.
\end{equation}
\end{remark}

\renewcommand{\theequation}{\thesection.\arabic{equation}}
\section{$1$-RIP for Expanders}
\label{app:RIP}

We present a simple argument that the perturbed matrix $\tilde{\A}$ has the $1$-RIP property if the underlying graph is an $(k,\epsilon)$, for $\epsilon<1/2$.
The $1$-RIP property states that for every $k$-sparse vector $\x$ and suitable constants $c_1,c_2$, the $\ell_1$ norm of $\|\A \x\|_1$ is close to the norm of $\x$:
\beq
(1-c_1) \|\x\|_1 \leq \|\A \x\|_1 \leq (1+c_2) \|\x\|_1.
\eeq
Berinde et al.~\cite{Indyk RIP} already show that adjacency matrices of expander graphs will have this property, also for $p$ norms where $p \leq 1-1/\log n$.
The argument we present here also requires $\epsilon<1/2$, but is arguably simpler and easily extends to the perturbed case.

Consider $\tilde{\A}$ to be the perturbed adjacency matrix of a $(k,\epsilon)$ unbalanced expander for $\epsilon<1/2$ and each nonzero entry is in $[1-\epsilon_1, 1+\epsilon_1]$.
Consider $S$, the support set of $\x$. By Hall's theorem since every set $S$ of size of size up to $k$ has $d(1-\epsilon)|S|$ neighbors, there must exist a $d(1-\epsilon)$-matching, i.e. every node in $S$ can be matched to $d(1-\epsilon)$ left nodes. Decompose the measurement matrix
\beq
\tilde{\A}=\A_M + \A_C.
\eeq
Where $\A_M$ is supported on the $d(1-\epsilon)$-matching (i.e every row has one non-zero entry and every column has $d(1-\epsilon)$ non-zero entries).
The remainder matrix $\A_C$ has $\epsilon d$ non-zero entries in each column, and notice that the decomposition is adapted to the support of the vector $\x$.
By the triangle inequality:
\beq
\|\A \x\|_1 \geq \|\A_M \x\|_1 - \|\A_C \x\|_1.
\eeq

It is easy to see that
\beq
\|\A_M \x\|_1 \geq (1-\epsilon_1) d (1-\epsilon) \|\x\|_1,
\eeq since $\A_M \x$ is a vector that contains $d(1-\epsilon)$ copies of each entry of $\x$ multiplied by coefficients in $[1-\epsilon_1, 1+\epsilon_1]$.
Also since each column of $\|\A_C\|_1$ contains $\epsilon d$ non-zero entries,
\beq
\|\A_C \x\|_1 \leq (1+\epsilon_1) \epsilon d \|\x\|_1,
\eeq
since each entry of $\A_M \x$ is a summation of terms of $\x$ and $\|\A_M \x\|$ is also a summation in which each entry of $\x$ appears $d\epsilon$ times, multiplied by coefficients in $[1-\epsilon_1, 1+\epsilon_1]$.
The same argument yields the upper bound:
\beq
\|\A \x\|_1 \leq d \|\x\|_1
\eeq

Therefore, putting these together we obtain:
\beq
(1- \epsilon_1 -2 \epsilon) \|\x\|_1 \leq
\|\A \x\|_1 \leq d \|\x\|_1.
\eeq
Therefore we need an expander with $\epsilon \geq (1-\epsilon_1)/2$ which for arbitrarily small perturbations goes to the $\epsilon<1/2$ limit.

\renewcommand{\theequation}{\thesection.\arabic{equation}}
\section{Proof of Robustness of Algorithm \ref{algo: noisy case}}
\label{app:noise}

We first state the following lemma from ~\cite{Indyk RIP}:
\begin{lem}{ Consequence  of Lemma 9 of ~\cite{Indyk RIP}:}\label{lem: RIP indyk}
If A is the adjacency matrix of a ($k$,$\epsilon$)-expander with $\epsilon < 0.5$
and $\u$ is a $k$-sparse vector, then  $d(1-2\epsilon)\|\u\|_1\leq \|\A\u\|\leq d\|\u\|_1$.
\end{lem}

We have presented a new proof of this Lemma in the Appendix \ref{app:RIP}, which is based on Generalized Hall's Theorem. That proof allows us to state with certainty that with suitable perturbations, $\tilde{\A}$ will also have a $1$-RIP.

By rearranging the rows and columns of $\A$, we may assume $\x=\left[\begin{array}{c}
      \x_1 \\ \x_2 \end{array}\right]$, $\y=\left[\begin{array}{c}
      \y_1 \\ \y_2 \end{array}\right]$, $\v=\left[\begin{array}{c}
      \v_1 \\ \v_2 \end{array}\right]$   and $\A=\left[\begin{array}{cc}
\A_{11} & 0 \\
\A_{12} & \A_2
\end{array}\right]$, where $\y_1$ and $\y_2$ are those obtained by the algorithm, $\x_1=\x_{S_1}$ and $\x_2=\x_{S_2}$. Also let $\e=\x-\hat{\x}$ be the reconstruction error vector. By (\ref{eq: noisy obs.}) we then have
\begin{eqnarray}
\y_1=\A_{11}\x_1 +\v_1 \\ \label{eq: y1=...}
\y_2=\A_{12}\x_1 + \A_{2}\x_2 + \v_2 \\ \label{eq: y2=...}
\e =\left[\begin{array}{c}
      \x_1 \\ \x_2-\hat{\x}_2 \end{array}\right]
\end{eqnarray}

Hence we have:
\begin{eqnarray}\label{eq: bound e1}
\|\x_1\|_1 \leq \|\A_{11}\x_1\|_1 = \|\y_1-\v_1\|_1 \leq\|\y_1\|_1+\|\v_1\|_1 \leq 2\|\v_1\|_1
\end{eqnarray}
The first inequality holds as a result of nonnegativity of $\x_1$ and $\A_{11}$, and the fact that every column of $\A_{11}$
has at least $1$ nonzero entry. The last inequality is a consequence of the choice of $\y_1$ in step $1$ of the algorithm and the fact that $\A\x$ is $m-rd$ sparse. Let's assume $\y_2 = \A_2\x_2 + \ddelta_2$. From the way $\hat{\x}_2$
is driven in step 4 of the algorithm it follows that:
\begin{equation}
\|\ddelta_2\|_1 \leq \|\A_{12}\x_1+\v_2\|_1
\end{equation}
And thus
\begin{equation}
\|\A_2(\x_2-\hat{\x}_2)\|_1 = \|\ddelta_2-\A_{12}\x_1-\v_2\|_1 \leq 2\|\A_{12}\x_1+\v_2\|_1 \leq 2d\|\x_1\|_1+2\|\v_2\|_1
\end{equation}
Using this along with the $1$-RIP condition of Lemma \ref{lem: RIP indyk} for the sparse vector $\u=\left[\begin{array}{c}
      0\\ \x_2-\hat{\x}_2 \end{array}\right]$ we get:
\begin{equation}\label{eq: bound e2}
c_1\|\x_2-\hat{\x}_2\|_1\leq 2d\|\x_1\|_1+2\|\v_2\|_1
\end{equation}
Where $c_1=(1-2\epsilon)d$.
Equations (\ref{eq: bound e1}) and (\ref{eq: bound e2}) result in:
\begin{equation}
\|\e\|_1 \leq (2+\frac{4d}{c_1})\|\v_1\|_1+\frac{2}{c_1}\|\v_2\|_1\leq \frac{6-4\epsilon}{1-2\epsilon}\|\v\|_1
\end{equation}
Therefore we have been able to bound the $\ell_1$ norm of the error with a constant factor of $\ell_1$ norm of noise as desired.

\begin{remark}
As soon as a $p$-RIP can be proven to hold for $\A$, a similar statement bounding $\ell_p$ norm of error with that of noise can be authentically established. However step 4 of the algorithm needs to be revised and $\ell_1$ optimization should be replaced  with $\ell_p$. In particular $p=2$ is of practical interest and the $\ell_2$ optimization of step $4$ is equivalent to the pseudo-inverse multiplication of the noise-less algorithm. However, as we mentioned earlier $2$-RIP does not hold for $0$-$1$ matrices. We speculate that a suitable perturbation can force singular values of $\A$ to jump above a constant,
and thus gift $\A$ a $2$-RIP condition.

\end{remark}

\renewcommand{\theequation}{\thesection.\arabic{equation}}
\section{Elementary bounds on binomial coefficients}

\label{AppBin}
For each $\beta \in (0,1)$, define the binomial entropy $H(\beta)= - \beta \log_2 \beta - (1-\beta) \log_2 (1-\beta)$ (and $H(0) =
H(1) = 0$ by continuity).  We make use of the following standard
bounds on the binomial coefficients
\begin{equation}\label{eq: binomial bounds}
n \, \left[ H\left(\frac{k}{n}\right) - \frac{ \log_2 (n+1)}{n}
\right] \; \leq \; \log_2 {n \choose k} \; \leq \; n \, \left[
H\left(\frac{k}{n}\right) + \frac{ \log_2 (n+1)}{n} \right].
\end{equation}

\end{document}